\documentclass{JHEP3}

\def \be {\begin{equation}}
\def \ee {\end{equation}}
\def \bea {\begin{eqnarray}}
\def \eea {\end{eqnarray}}
\def \nn {\nonumber}

\def \a {\alpha}
\def \b {\beta}
\def \g {\gamma}

\def \d {\delta}

\def \m {\mu}
\def \n {\nu}
\def \k {\kappa}

\def \s {\sigma}
\def \r {\rho}
\def \o {\omega}
\def \O {\Omega}
\def \th {\theta}
\def \Th {\Theta}

\def \t {\tau}
\def \dag {\dagger}
\def \p {\partial}
\def\sY {\sum_I s^IY^I}
\def\bd{\begin{document}}
\def\ed{\end{document}}
\def\nn{\nonumber}
\def\bea{\begin{eqnarray}}
\def\eea{\end{eqnarray}}
\let\bm=\bibitem
\let\la=\label

\def\N{{\cal N}}
\def\sst{\scriptscriptstyle}
\def\thetabar{\bar\theta}
\def\Tr{{\rm Tr}}
\def\one{\mbox{1 \kern-.59em {\rm l}}}

\def\a{\alpha}      \def\da{{\dot\alpha}}
\def\b{\beta}       \def\db{{\dot\beta}}
\def\c{\gamma}  \def\C{\Gamma}  \def\cdt{\dot\gamma}
\def\d{\delta}  \def\D{\Delta}  \def\ddt{\dot\delta}
\def\e{\epsilon}        \def\vare{\varepsilon}
\def\f{\phi}    \def\F{\Phi}    \def\vvf{\f}
\def\h{\eta}
\def\k{\kappa}
\def\l{\lambda} \def\L{\Lambda}
\def\m{\mu} \def\n{\nu}
\def\o{\omega}
\def\P{\Pi}
\def\r{\rho}
\def\s{\sigma}  \def\S{\Sigma}
\def\t{\tau}
\def\th{\theta} \def\Th{\Theta} \def\vth{\vartheta}
\def\X{\Xeta}
\def\z{\zeta}
\def\w{\wedge}
\def\u{\underline}
\def\hs{\hspace}

\def\cA{{\cal A}} \def\cB{{\cal B}} \def\cC{{\cal C}}
\def\cD{{\cal D}} \def\cE{{\cal E}} \def\cF{{\cal F}}
\def\cG{{\cal G}} \def\cH{{\cal H}} \def\cI{{\cal I}}
\def\cJ{{\cal J}} \def\cK{{\cal K}} \def\cL{{\cal L}}
\def\cM{{\cal M}} \def\cN{{\cal N}} \def\cO{{\cal O}}
\def\cP{{\cal P}} \def\cQ{{\cal Q}} \def\cR{{\cal R}}
\def\cS{{\cal S}} \def\cT{{\cal T}} \def\cU{{\cal U}}
\def\cV{{\cal V}} \def\cW{{\cal W}} \def\cX{{\cal X}}
\def\cY{{\cal Y}} \def\cZ{{\cal Z}}

\def\ua{\underline{\alpha}} \def\ubb{\underline{\beta}}
\def\ug{\underline{\gamma}}
\def\ub{\underline{\phantom{\alpha}}\!\!\!\beta}
\def\uc{\underline{\phantom{\alpha}}\!\!\!\gamma}
\def\um{\underline{\mu}} \def\un{\underline{\nu}}
\def\ud{\underline\delta}
\def\ue{\underline\epsilon}
\def\una{\underline a}\def\unA{\underline A}
\def\unb{\underline b}\def\unB{\underline B}
\def\unc{\underline c}\def\unC{\underline C}
\def\und{\underline d}\def\unD{\underline D}
\def\une{\underline e}\def\unE{\underline E}
\def\unf{\underline{\phantom{e}}\!\!\!\! f}\def\unF{\underline F}
\def\unm{\underline m}\def\unM{\underline M}
\def\unn{\underline n}\def\unN{\underline N}
\def\unp{\underline{\phantom{a}}\!\!\! p}\def\unP{\underline P}
\def\unq{\underline{\phantom{a}}\!\!\! q}
\def\unQ{\underline{\phantom{A}}\!\!\!\! Q}
\def\unH{\underline{H}}

\def\As {{A \hspace{-6.4pt} \slash}\;}
\def\bs {{b \hspace{-6.4pt} \slash}\;}
\def\Ds {{D \hspace{-6.4pt} \slash}\;}
\def\ds {{\del \hspace{-6.4pt} \slash}\;}
\def\ss {{\s \hspace{-6.4pt} \slash}\;}
\def\ks {{ k \hspace{-6.4pt} \slash}\;}
\def\ps {{p \hspace{-6.4pt} \slash}\;}
\def\pas {{{p_1} \hspace{-6.4pt} \slash}\;}
\def\pbs {{{p_2} \hspace{-6.4pt} \slash}\;}

\def\Fh{\hat{F}}
\def\Vh{\hat{V}}
\def\Xh{\hat{X}}
\def\ah{\hat{a}}
\def\xh{\hat{x}}
\def\yh{\hat{y}}
\def\ph{\hat{p}}
\def\xih{\hat{\xi}}

\def\psit{\tilde{\psi}}
\def\Psit{\tilde{\Psi}}
\def\tht{\tilde{\th}}

\def\At{\tilde{A}}
\def\Qt{\tilde{Q}}
\def\Rt{\tilde{R}}
\def\Nt{\tilde{N}}

\def\at{\tilde{a}}
\def\st{\tilde{s}}
\def\ft{\tilde{f}}
\def\pt{\tilde{p}}
\def\qt{\tilde{q}}
\def\vt{\tilde{v}}
\def\nt{\tilde{n}}

\def\delb{\bar{\partial}}
\def\bz{\bar{z}}
\def\bD{\bar{D}}
\def\bB{\bar{B}}

\def\bk{{\bf k}}
\def\bl{{\bf l}}
\def\bp{{\bf p}}
\def\bq{{\bf q}}
\def\br{{\bf r}}
\def\bx{{\bf x}}
\def\by{{\bf y}}
\def\bR{{\bf R}}
\def\bV{{\bf V}}

\def\d{\delta}\def\D{\Delta}\def\ddt{\dot\delta}

\def\p{\partial} \def\del{\partial}
\def\xx{\times}
\def\uno{\mbox{1 \kern-.59em {\rm l}}}

\def\trp{^{\top}}
\def\inv{^{-1}}
\def\dag{{^{\dagger}}}
\def\pr{^{\prime}}

\def\rar{\rightarrow}
\def\lar{\leftarrow}
\def\lrar{\leftrightarrow}

\title{Operator product expansion of Wilson surfaces from M5-branes}

\author{Bin Chen\\
Department of Physics,\\
Peking University,\\
Beijing 100871, P.R. China\\
and\\ KITPC, CAS\\
Beijing 100080, P. R. China\\ \email{bchen01@pku.edu.cn}}

\author{Chang-Yong Liu\\
Institute of Theoretical Physics\\
 Chinese Academy of Science,\\
 Beijing 100080, P.R. China\\
 and\\
 Graduate University of Chinese Academy of Science,\\
 Beijing 100080, P.R. China\\
\email{lcy@itp.ac.cn}}

\author{Jun-Bao Wu\\International School for Advanced Studies (SISSA) and INFN,\\
via Beirut 2-4, I-34014 Trieste, Italy\\
\email{wujunbao@sissa.it}}

\date{\today}

\abstract{The operator product expansion (OPE) of the Wilson surface
operators in six-dimensional (2, 0) superconformal field theory is
studied from AdS/CFT correspondence in this paper. We compute the
OPE coefficients of the chiral primary operators using the M5-brane
description for spherical Wilson surface operators in higher
dimensional representations. We use the non-chiral M5-brane action
in our calculation. We also discuss their membrane limit, and
compare our results with the ones obtained from membrane
description. }

\preprint{\ CAS-KITPC/ITP-018\\ SISSA-83/2007/EP}

\newpage
\begin{document}

\section{Introduction}

Supersymmetric Wilson loops play an important role in
$AdS_5/CFT_4$ correspondence \cite{Mal97, Gubser:1998bc,
Witten:1998qj, Aharony:1999ti}. On the field theory side, the
calculation of the expectation values of half-BPS circular Wilson loops
could be reduced to the corresponding calculation in a
zero-dimensional matrix model \cite{Zarembo}. The reduction to the
matrix model relies on the fact that the perturbative
contributions to the expectation value are believed to be only
from the rainbow graphs in Feynman gauge
\cite{Zarembo} and this is confirmed by using the conformal transformation
which links the straight Wilson line and the circular Wilson loop \cite{Drukker01}. 
It is remarkable that the computations using the matrix model give
us the results to all orders of $g_{YM}^2N$ and to all orders of
$1/N$. The dependence on $1/N$ indicates that in order to have a
good dual description of these BPS Wilson loops, one has to go
beyond the free string limit and consider the string interaction
on the $AdS_5$ side.

The original $AdS_5/CFT_4$ dictionary tells us that the dual
description of the Wilson loops in $AdS_5$ should be the
fundamental strings whose worldsheet boundaries are just the paths
used to define the Wilson loops in ${\cal N}=4$ Super-Yang-Mills
theory\cite{Rey:1998ik, Mal98}. The on-shell classical actions of
the strings give the expectation values of the Wilson loops, after
correctly including the boundary terms \cite{Gross}. However, the
field theory result indicates that this should not be the full
story and one should go beyond the free string limit. Later on,
people found that a better description for the half-BPS Wilson
loop in high rank representation of gauge group is using
D3-brane and/or D5-brane configurations\cite{Drukker,
Yamaguchi:2006D5, Gomis:2006sb, Gomis:2006im}. The D3-brane
configuration gives a good description for the Wilson loops in the
symmetric representation, while the D5-brane gives a good
description for the ones in the antisymmetric representation. The
original string picture is a good description only for the Wilson
loops in the fundamental representation or low dimensional
representations. The D-brane description of the Wilson loops in
high dimensional representations can be understood as dielectric
effect\cite{Myers:1999, Rodriguez2006}: due to the interaction
among many coincide fundamental strings in the self-dual RR
background, the strings blow up to higher dimensional D-branes.
The expectation values of the Wilson loops can be computed from
the action of the classical D-brane solutions in the large N
limit, appropriately taken into account of the boundary terms. The
computations using D-branes successfully reproduce the all-genus
results from matrix model calculation\cite{Drukker,
Yamaguchi:2006D5}. Furthermore D3-brane description of some
1/4-BPS Wilson loops was given in \cite{Drukker:2006zk} and the
D-brane description of 1/2-BPS Wilson-'t Hooft operators was given
in \cite{ChenHe}. Some further studies of higher rank Wilson loops
using matrix model can be found in \cite{Okuyama:2006jc, Hartnoll:2006is}.

Another interesting issue on Wilson loops is to calculate their
OPE.  When we probe the Wilson loop from a distance much larger
than the size of the loop, this Wilson loop operator can be
expanded as a linear combination of local operators. When the
local operator is a chiral primary operator, the OPE coefficient
can be computed either from the correlation function of two Wilson
loops or from the correlation of the Wilson loop with this
operator \cite{Corrado}. According to AdS/CFT correspondence, in
the large N and large $g_{YM}^2N$ limit this OPE coefficient can
be computed from the coupling to the string worldsheet
corresponding to the Wilson loop of the supergravity mode
corresponding to the chiral primary operator \cite{Corrado}. When
the Wilson loop operator is in high dimensional representation,
the OPE coefficients can be computed from the coupling to the
corresponding D-branes of the supergravity modes
\cite{Giombi:2006de}.

Motivated by the success in the Wilson loop case, we would like to
consider its cousin in six-dimensional field theory in the
framework of $AdS_7/CFT_6$ correspondence. Here $CFT_6$ is a
six-dimensional superconformal field theory with $(2,
0)$-supersymmetries. Its field content is of a tensor multiplet,
including a 2-form $B_{\mu\nu}$, four fermions and five scalars;
the field strength of this 2-form is (anti)-self-dual. The strong
version of the $AdS_7/CFT_6$ correspondence claims that this field
theory is dual to the M-theory on the background $AdS_7 \times
S^4$. This correspondence was obtained by considering $N$
coinciding M5-branes in M-theory.
The low energy limit of the worldvolume theory is the above
six-dimensional $A_{N-1}, (2, 0)$ superconformal field theory
\cite{Strominger:1995ac, Witten95}.  The near horizon limit of the
supergravity solution corresponding to these M5-branes will give
$AdS_7\times S^4$ background with 4-form flux. Similar to the
$AdS_5/CFT_4$ case, this near horizon limit led Maldacena to
propose the above correspondence \cite{Mal97}.\footnote{In
\cite{Witten:1998zw}, this correspondence was used to study the
nonsupersymmtric QCD in four dimensions.} Unfortunately, unlike
the well-studied $AdS_5/CFT_4$ case, the $AdS_7/CFT_6$
correspondence is poorly investigated, although its study could be
essential for us to understand M-theory. The main obstacle is our
ignorance of the mysterious superconformal field theory. Due to
the existence of self-dual chiral 2-form, there is no lagrangian
formulation of the theory, even though the chiral theory is still
a local interacting field theory\cite{Witten96}. The theory has
been suggested to be described by DLCQ matrix
theory\cite{Seiberg97, aharony}. In any sense, it has not been
well understood. The AdS/CFT correspondence supplies a new tool to
probe this nontrivial six-dimensional field theory. The weak
version of the correspondence says that the large $N$ limit of the
$(2,0)$ field theory is dual to 11D supergravity on $AdS_7 \times
S^4$\cite{Mal97}. The chiral primary operators and the
corresponding supergravity modes in this case were studied in
\cite{AdS7CFT}. Some correlation functions of local operators were
computed in \cite{Corrado99}. These local operators were also studied
using M5-brane action\cite{Nurmagambetov:2001ab}.

In this six-dimensional superconformal field theory, the natural
cousin of Wilson loop operator is Wilson surface operator, a
non-local operator of dimension two. This operator could be
formally defined as \cite{Ganor}
 \be
 W_0(\Sigma)=\exp i\int_\Sigma B^+.
 \ee
Here $\Sigma$ is a two-dimensional surface. From AdS/CFT
correspondence, the Wilson surface operator should correspond to a
membrane ending on the boundary of AdS space\cite{Mal98, Corrado}.
Inspired by the D-brane description of Wilson loops in higher
dimensional representations, M5-brane description of the half-BPS
Wilson surface operators in higher dimensional representations were
studied in details in \cite{Chen}.\footnote{Similar brane
configurations for straight Wilson surface are discussed in
\cite{Lunin07} in Pasti-Sorokin-Tonin (PST) formalism as well. The
self-dual string soliton in $AdS_4\times S^7$ spacetime is
discussed in \cite{Lunin07,Chen2}.} The corresponding M5-brane
solutions of the covariant equations of motion have been found.
Both the straight Wilson surfaces and the spherical Wilson
surfaces were studied in this framework. For each case, two kinds
of solution were discovered. Both of them have worldvolume of
topology $AdS_3\times S^3$. The $AdS_3$ part is always in $AdS_7$,
while the $S^3$ part can be either in $AdS_7$ or in $S^4$.
Analogizing
 the D-brane description of the Wilson loops, we expect the first case
describe the Wilson surface in the symmetric representation, while the
second solution describe the Wilson surface in the anti-symmetric
representation. The expectation value of the Wilson surface should
be given by the action of the membrane or the M5-brane. Both
actions are divergent\cite{Corrado, Chen}. For the straight Wilson
surface, there is only quadratic divergence, but for the spherical
Wilson surface, there are both quadratic and logarithmic
divergences. The logarithmic divergence comes from conformal
anomaly of the surface operator\cite{Witten99}. The existence of
logarithmic divergence indicates that the expectation value of
Wilson surface may not be well-defined. Despite of this fact, the
OPE coefficients of the Wilson surface operators are still
well-defined. In \cite{Corrado99}, the OPE coefficients of the
chiral primary operators are computed using the membrane solution
found in \cite{Corrado}. The strategy is similar to the Wilson
loop case: one may treat the membrane as the source for the
supergravity fields in the bulk. The OPE coefficients could be
read off from the coupling to the membrane of the bulk
supergravity modes corresponding to the chiral primary operators.

The main subject of this paper is to compute the OPE coefficients
for the Wilson surface operator in higher dimensional
representation using the M5-brane solutions mentioned above. We
compute these OPE coefficients from the correlation functions of
the Wilson surface with the chiral primary operators. Instead of
taking the membrane as source, we take the M5-brane as the source
and study its response to the bulk gravity modes. Unlike the cases
of D-brane and M2-brane, the dynamics of M5-brane is much more
subtler. Various actions of M5-branes are given in
\cite{Sundell97}-\cite{Sezgin99}. In this paper we will use the
non-chiral action in \cite{Sezgin99} to compute the OPE
coefficients. The virtue of this action is that we need not to
introduce any auxiliary fields.

The paper is organized as follows. In section 2, we review the
computations of the OPE coefficients using membrane solution.
Section 3 is devoted to a very brief review of the non-chiral
action of M5-brane. The computations using M5-brane solution is
present in the following two sections. Section 4 is for the
symmetric case and section 5 is for the anti-symmetric case. We
end with the conclusion and discussions. We put the technical
details about the variation of dual six-form gauge potential $\d
C_6$ in the appendix.

\section{Review of the OPE of the Wilson surface in the fundamental presentation}

In this paper, we only consider the spherical Wilson surface
operators. When we probe the Wilson surface from a distance quite
larger than its radius $r$, the operator product expansion of the
Wilson surface operators could be: \be W(S)=<W(S)>(1+\sum_{i,
n}c_i^n r^{\Delta^n_i}{\cal O}_i^n), \ee where ${\cal O}_i^n$ are
operators with conformal weights $\Delta^n_i$. Here we use ${\cal
O}_i^0$ to denote the $i$-th primary field and ${\cal O}_i^n$ for
$n>0$ to denote its conformal descendants. The OPE coefficients of
the chiral primary operator ${\cal O}_i^0$ can be obtained from
the $r/L$ expansion of the correlation function of the Wilson
surface with this chiral primary operator, \be {<W(S){\cal
O}^0_i>\over <W(S)>}=c^0_i {r^{\D^0_i}\over
L^{2\D^0_i}}+\sum_{m>0}c^m_ir^{\D^m_i} <{\cal O}^m_i{\cal O}^0_i>,
\ee where $L$ is the distance from the Wilson surface to the local
operator, and we have assume that the local operators have been
normalized.

These operators should be bosonic and $S_{N}$ symmetric, since
they should have the same symmetry property of the Wilson surface.
Based on the experience from the supersymmetric Wilson loops in
${\cal N}=4$ SYM, the half-BPS Wilson surface should also coupled
to the five scalars. This coupling is determined by a vector
$\tilde\theta^I(s)$ in $S^4$ \cite{Corrado}. We consider the case
when $\tilde\theta^I(s)=\tilde\theta^I$ is a constant, i. e. a fixed
point in $S^4$. Then the R-symmetry group is broken from $SO(5)$
to $SO(4)$. The local operators which appear in the OPE of the
Wilson surface should also be in the representation of $SO(5)$
whose decomposition includes singlet of $SO(4)$. In
this paper, we will compute the OPE coefficient of the operator
${\cal O}_{\Delta}$ in the rank $k$ symmetric, traceless
representation of $SO(5)$. This operator satisfy the above
constraints and is a chiral primary operator of dimension
$\Delta=2k$ \cite{aharony}. The dimension of this operator is
protected by supersymmetries.

\subsection{Review of the corresponding supergravity modes}

 In the following, we would like to review the
supergravity modes corresponding to this chiral primary operators.
To do this, we would like to first review the $AdS^7\times S^4$
solution of 11d supergravity. This solution is maximally
supersymmetric.

The bosonic equations of motion of 11d supergravity are \footnote{We use the following notation: $m, n, \cdots$ refer to the
coordinate indices of $AdS_7\times S^4$, $\mu, \nu, \cdots$ refer to the coordinate indices of the $AdS_7$ part, $\alpha, \beta, \cdots$
refer to ones of the $S^4$ part, and the underline indices refer to target space ones.}:
 \bea
 R_{\underline{m}\underline{n}}&=&\frac{1}{2\xx 3!}H_{\underline{m}\underline{p}\underline{q}\underline{r}}
 H_{\underline{n}}^{~\underline{p}\underline{q}\underline{r}}-\frac{1}{6\xx
 4!}g_{\underline{m}\underline{n}}H_{\underline{p}\underline{q}\underline{r}\underline{s}}H^{\underline{p}\underline{q}\underline{r}\underline{s}}, \\
 0&=&\p_{\underline{m}}\left(\sqrt{-g}H^{\underline{m}\underline{n}\underline{p}\underline{q}}\right)+\frac{1}{2\xx
 (4!)^2}\epsilon^{\underline{m}_1\cdots \underline{m}_8\underline{n}\underline{p}\underline{q}}H_{\underline{m}_1\cdots \underline{m}_4}
 H_{\underline{m}_5\cdots \underline{m}_8}.\eea
And the metric and background 4-form flux of $AdS_7\times S^4$ are
 \bea
 ds^2&=&\frac{1}{y^2}(dy^2-dt^2+dx^2+dr^2+r^2d\Omega_3^2)+\frac{1}{4}d\Omega_4^2\nn\\
 H_4&=&\frac{3}{8}\sin^3\z_1\sin^2\z_2\sin\z_3d\z_1\w d\z_2\w
 d\z_3\w d\z_4
 \eea
 where $d\O_3^2$ is the metric of unit $S^3$ and $d\O_4^2$ is the
 metric of unit $S^4$. The 4-form field strength fills in $S^4$,
 and $\z_i\,(i=1,2,3, 4)$ are the angular coordinates in $S^4$.
 We have rescaled the radius of $AdS_7$ to be $1$, then the radius
 of $S^4$ is $1/2$. From the
 $AdS_7/CFT_6$ duality, we know that\begin{equation}
 l_p=(8\pi N)^{-\frac{1}{3}},\end{equation}
  where $l_p$ is the Planck constant in eleven dimension.
The 4-form field strength $H_4$ and its Hodge dual 7-form field
strength $H_7$ are related to the corresponding gauge potentials
$C_3$ and
$C_6$ by \bea H_4&=&dC_3, \nn\\
H_7&=&dC_6+{1\over 2}C_3\w H_4. \label{flux} \eea

  Now we consider the fluctuation around the above background to get the states of
  11d supergravity in this background \cite{Nieuwenhuizen1, Nieuwenhuizen2, Nieuwenhuizen3}.
 We can decompose the fluctuated metric as
 \be G_{\underline{m}\underline{n}}=g_{\underline{m}\underline{n}}+h_{\underline{m}\underline{n}}, \ee where
 $g_{\underline{m}\underline{n}}$ is the background metric, $h_{\underline{m}\underline{n}}$ is the
 fluctuations. The fluctuation of the three form gauge potential is
 \begin{equation}
\delta C_{\underline{m}\underline{n}\underline{p}}=a_{\underline{m} \underline{n}\underline{p}}
\end{equation}

 We first decompose $h_{\underline{\alpha}\underline{\beta}}$ into the trace part and the
 traceless part:
 \be h_{\underline{\alpha}\underline{\beta}}=h_{(\underline{\alpha}\underline{\beta})}+{1\over 4}h_2 g_{\underline{\alpha}\underline{\beta}}. \ee
  Then we decompose $h_{\um\un}$ as
 \be h_{\um\un}=h^\prime_{(\um\un)}+\left({h^\prime \over 7}-{h_2\over 5}\right)g_{\um\un}. \ee
 Here $(\underline{m}\underline{n})$ indicates that we take the symmetric traceless part.

 In the gauge defined by
 \be \nabla^{\ua} h_{(\ua\ubb)}=\nabla^{\ua} h_{\ua\um}=\nabla^{\ua} a_{\ua \underline{m}\underline{n}}=0, \ee
 $h^\prime, h_2, h_{(\ua\ubb)}, h^\prime_{(\um\un)}$ and $a_{\underline{m}_1\underline{m}_2\underline{m}_3}$ have the following expansion:
 \bea
 h^\prime=\sum_I h^{\prime I}Y^I, & & h_2=\sum_I h^I_2 Y^I, \nn \\
 h_{(\ua\ubb)}=\sum_I\phi^IY^I_{(\ua\ubb)}, & & h^\prime_{(\um\un)}=\sum_I h^{\prime I}_{(\um\un)} Y^I,
 \eea
 and
  \begin{equation}
a_{\ua \ubb \ug}=\sum_{I} 6\sqrt{2}\epsilon_{\ua \ubb \ug \ud}b^I
\nabla^{\ud} Y^I.
\end{equation}

 Here $Y^I$ and $Y^I_{(\a\b)}$ are scalar and rank 2, symmetric traceless tensor harmonics
on four-sphere with radius $1/2$,  respectively.
 They satisfy the following equations
 \be
\nabla^{\ua} \nabla_{\ua} Y^I=-4k(k+3)Y^I, \ee and \be
\nabla^{\ua} \nabla_{\ua}
Y^I_{(\ubb\ug)}=-4[k(k+3)-2]Y^I_{(\ubb\ug)}, \ee
respectively\footnote{We use the same
normalization of the harmonic functions as in \cite{Corrado99}.}.
The index $I$ is the abbreviation of $(l_4, \cdots, l_1)$ which
satisfy \be l_4\equiv k\geq l_3 \geq l_2 \geq |l_1|.\ee

Using the above expansions, we can obtain the linearized equations
of motion which we will not repeat here. The modes $h_2$ and $b$
satisfy a set of coupled equations of motion. The mass
eighenvectors and eighenvalues are \bea
s^I&=&{k\over 2k+3}[h^I_2+32\sqrt{2}(k+3)b^I], \hspace{3ex}m_s^2=4k(k-3),\hspace{3ex} k\ge 2,\\
t^I&=&{k+3\over 2k+3}[h^I_2-32\sqrt{2}kb^I],\hspace{3ex}
m_t^2=4(k+3)(k+6),\hspace{3ex} k\ge 0. \eea $s^I$ transforms in
the same representation of the R-symmetry group $SO(5)$ as ${\cal
O}_\Delta$, and it is the supergravity mode corresponding to
${\cal O}_\Delta$ \cite{AdS7CFT}.

Since we are only interested in the OPE coefficients of ${\cal
O}_\D$, we can set the other modes to be zero. From $t^I=0$, we
get \be h^I_2=32\sqrt{2}kb^I, \ee so \be s^I=32\sqrt{2}kb^I=h^I_2.
\ee

Using the results in \cite{Corrado99}, we can express the
fluctuation of the background in terms of $s^I$ as:
 \be h^I_{\ua\ubb}={1\over 4}g_{\ua\ubb}s^I,\ee

 \begin{equation}
h^I_{\um\un}={3\over 16k(2k+1)}\nabla_{(\um}\nabla_{\un)}
s^I-{1\over 14}g_{\um\un}s^I,
\end{equation}
and
\begin{equation}
\delta C_{\ua \ubb \ug}=\sum_{I}{3\over 16k }\epsilon_{\ua \ubb
\ug \ud}s^I \nabla^{\ud} Y^I.
\end{equation}

\subsection{Review of the computations of the OPE coefficients}

In this subsection we will review the membrane solution
corresponding to the Wilson surface in the fundament representation in
\cite{Corrado} and the computations of the OPE coefficients using
this solution \cite{Corrado99}.

The membrane solution can be described more conveniently in the
Euclidean version of $AdS_7$ space and using the Poincar\'e
coordinates. In this coordinate system, the metric of the $AdS_7$
space is \be ds^2={1\over y^2}(dy^2+\sum_{i=1}^6dx_i^2). \ee

Consider a spherical Wilson surface with radius $r$ described by \be
x_1^2+x_2^2+x_3^2=r^2, \ee in the boundary of $AdS_7$. The membrane
solution corresponding to this Wilson surface can be parametrized as
following: \bea
x_1&=&\sqrt{r^2-y^2}\cos\theta,\nn\\
x_2&=&\sqrt{r^2-y^2}\sin\theta\cos\psi,\nn\\
x_3&=&\sqrt{r^2-y^2}\sin\theta\sin\psi, \eea where $0\le y\le r,
0\le\theta\le \pi, 0\le\psi\le 2\pi$.

To compute the OPE coefficient from this membrane solutions, we
need to use the action of the M2-brane. The bosonic part of this
action is \cite{Bergshoeff:1987cm} \be S_{M2}=T_2\int
(d\mbox{Vol}-\underline{C}_3), \ee where $T_2$ is the tension of
M2-brane: \be T_2={1\over (2\pi)^2l_p^3}={2N \over
\pi},\label{eqt2} \ee and $\underline{C}_3$ is the pullback of the
bulk 3-form gauge potential to the worldvolume of the
membrane\footnote{In this paper, we use the underline indices to
denote the target space indices. We also use the underline to
denote the pullback of bulk gauge potential or field strength to
the worldvolume of M2-brane or M5-brane. We hope that this will
not produce confusion. }. Since the worldvolume of the membrane is
completely embedded in the $AdS_7$ part of the background, so the
pullback of $\delta C_3$ to the membrane worldvolume is zero. Then
the only contribution is from the Nambu-Goto part of the action:
\be \delta S_{M2}={1\over 2}T_2\int d\mbox{Vol} g^{mn} h_{mn}. \ee

After we compute the fluctuation of the action due to the
supergravity modes, we write $s^I$ as $s^I(\vec{x}, y)=\int
d^6\vec{x}^\prime G_\Delta(\vec{x}^\prime; \vec{x},
y)s^I_0(\vec{x}^\prime)$. Here \be G_\Delta(\vec{x}^\prime;
\vec{x}, y)=c \left(y\over
y^2+|\vec{x}-\vec{x}^\prime|^2\right)^\Delta,\label{green}\ee is
the bulk-to-boundary propagator and c is the following constant:
\be c={8^{2+k}(2k-3)(2k-1)(2k+1)\Gamma(k+3/2)\over
9\pi^{1/2}N^3\Gamma(k)}. \label{eqc}\ee Then the correlation
function we needed to compute is: \be {\langle W({\cal S}, L)
{\cal O}_\Delta(0) \rangle\over \langle W({\cal S}, L) \rangle}
\sim -{1\over {\cal N}^I}{\d S_{M2}\over \d
s^I_0(\vec{x})}.\label{opem2}\ee Here \be{\cal
N}^I=-2^{3k/2+3}{(2k-3)(2k+1)\over
3\pi^{1/4}N^{3/2}}\sqrt{(2k-1)\Gamma(k+1/2)\over
\Gamma(k)}\label{eqni}\ee is used to set the normalization of the
operator and this constant is fixed by requiring the coefficient
of the 2-point function to be unit.

 Since we
only need to compute this function to the first order of $r/L$, we
can use the following approximation for the bulk-to-boundary
propagator\cite{Corrado}: \be G_\Delta(\vec{x}^\prime; \vec{x},
y)\simeq c {y^\Delta \over L^{2\Delta} }.\ee From
eq.~(\ref{green}), we find that to the first order of $r/L$, we
have the following approximations:

\begin{equation}
\partial_{\um} s^I\simeq \delta^y_{\um}{\Delta \over
y}s^I,\hspace{3ex}
\partial_{\um} \partial_{\un} s^I\simeq \delta^y_{\um}\delta^y_{\un}{\Delta(\Delta-1) \over y^2}s^I,\label{eqy}\end{equation}
By using this and
\begin{equation}
\Gamma^{y}_{\um\un}=yg_{\um\un}-{2\over
y}\delta^y_{\um}\delta^y_{\un}
\end{equation}
in Poincar\'e coordinate, we get
\begin{equation}
h^I_{\um\un}\simeq-{1\over 8}g_{\um\un}s^I+{3\over
8}\delta^y_{\um}\delta^y_{\un} {1\over y^2}s^I.
\end{equation}
From this, we have \be \d S_{M2}=-{3T_2\over 16}\int d
\mbox{Vol}{y^2\over r^2}\sY(\tilde\theta). \ee

By using eq.~(\ref{opem2}), (\ref{eqc}), (\ref{eqni}),
(\ref{eqt2}), we get \be {\langle W({\cal S}, L) {\cal
O}_\Delta(0) \rangle\over \langle W({\cal S}, L) \rangle} \sim
-2^{(3k+1)/2}\pi^{1/4}\sqrt{\Gamma(k)\over N\Gamma(k-{1\over
2})}{r^\Delta \over L^{2\Delta}} Y^I(\tilde\theta).\ee So the OPE
coefficients are\footnote{Notice that $Y^I$ is not included in the
OPE coefficients.} \be c_{\mbox{\small fumd.},
\Delta}=-2^{(3\Delta+2)/4}\pi^{1/4}\sqrt{\Gamma(\Delta/2)\over
N\Gamma((\Delta-1)/2)}.\label{eqcf}\ee

\section{The non-chiral action of M5-brane}

Compared to D-branes, the action of M5-brane is more involved.
Various actions of M5-branes were given in
\cite{Sundell97}-\cite{Sezgin99}. Different choice of action gives
equivalent equations of motion \cite{Sorokin97,
Sezgin99}. In this paper we will use the non-chiral action in
\cite{Sezgin99} to compute the OPE coefficients. There is a 3-form
field strength $H_3$ on the worldvolume of the M5-brane. This
field strength is related to a 2-form potential $A_2$ by \be
H_3=dA_2-\underline{C}_3,\label{eqh3} \ee so $H_3$ satisfies the
following Bianchi identity: \be dH_3=-\underline{H}_4 \ee Here
$\underline{C}_3$ and $\underline{H}_4$ are the pull-back of
target space 3-form potential and 4-form field strength,
respectively.

The non-chiral action is given by
 \be\label{nc}
 S=S_{M5}-S_{WZ}=T_5\int (\frac{1}{2}\star {\cal K}-Z_6),
 \ee
 where
  \be
 {\cal
 K}=2\sqrt{1+\frac{1}{12}H^2+\frac{1}{288}(H^2)^2-\frac{1}{96}H_{abc}H^{bcd}H_{def}H^{efa}},
 \ee
 \bea
 Z_6&=&\underline{C}_6-\frac{1}{2}\underline{C}_3\w H_3,
 \eea
and $T_5$ is the tension of the M5-brane:
 \be
 T_5=\frac{1}{(2\pi)^5l^6_p}=\frac{2N^2}{\pi^3}.\label{eqt5}
 \ee
Here $\underline{C}_6$ is the pull-back of target space 6-form
potential.  The equations of motion are obtained from the variation
of the action with respect to the embedding $z^{\underline{m}}$ and the gauge
potential $A_2$. The equation of motion for 2-form potential is
equivalent to the Bianchi identity. In addition, one have to impose
the following non-linear self-duality condition \cite{Sezgin99}
 \be\label{selfdual}
 \ast H_3=\frac{\p {\cal K}}{\p H_3},
 \ee
by hand.

In the following two sections we will study the OPE of the Wilson
surface operators using this non-chiral action. For doing this, we
need to compute the variations of the action with respect to the
above fluctuations of the background fields reviewed in subsection
2.1. Since we only need to compute the fluctuation to the linear
order and the equations of motion are obtained from the variation
of the action with respect to $z^{\underline{m}}$ and $A_2$, we
can set the variations of $z^{\underline{m}}$ and $A_2$ to be
zero. Then from eq.~(\ref{eqh3}) we get $\d H_3=-\d
\underline{C}_3$.

\section{OPE of the Wilson surface in the symmetric representation}

In this section we will study the OPE of the Wilson surface
operator in the symmetric representation by using the M5-brane
solutions in \cite{Chen}. We would like to compute the OPE
coefficients of ${\cal O}_\Delta$ by compute the correlation
functions of the Wilson surface operator with ${\cal O}_\Delta$.
According to the AdS/CFT correspondence, we need to study the
coupling to this M5-brane of the corresponding supergravity modes
$s^I$.

\subsection{Review of the M5-brane solution}
First we would like to review the M5-brane solution corresponding
to the spherical Wilson surface operator in the symmetric
representation. As in \cite{Drukker}, it is more convenient to make a Wick
rotation in the $AdS_7$ space and choose the coordinates such that
the metric take the following form:
\begin{equation}
ds^2=\frac{1}{y^2}(dy^2+dr_1^2+r_1^2(d\alpha^2+\sin^2\alpha
d\beta^2)+dr_2^2+r_2^2(d\gamma^2+\sin^2\gamma d\delta^2)),
\label{metric7}
\end{equation}

The Wilson surface will be placed at $r_1=r$ and $r_2=0$. Let us
change the coordinates $(r_1,r_2,y)$ to $(\rho, \eta, \theta)$ by
the following transformation:
\begin{equation}
r_1=\frac{r\cos\eta}{\cosh\rho-\sinh\rho\cos\th},~~r_2=\frac{r\sinh\rho\sin\th}{\cosh\rho-\sinh\rho\cos\th}
~~,y=\frac{r\sin\eta}{\cosh\rho-\sinh\rho\cos\th}, \end{equation}
then we can rewrite the $AdS_7$ metric as
 \bea
 ds^2&=&\frac{1}{\sin^2\eta}\big(d\eta^2+\cos^2\eta(d\a^2+\sin^2\a
 d\b^2)+d\rho^2\nn \\ & &+\sinh^2\rho(d\th^2+\sin^2\th d\g^2+\sin^2\th
 \sin^2\g d\d^2)\big).\label{AdS7metric}
 \eea
 Here, the coordinates take the range
 $\rho\in[0,\infty),\th,\a,\g\in[0,\pi),\b,\d \in [0,2\pi),\eta
 \in [0,\pi/2)$.

The worldvolume of the M5-brane has topology $AdS_3\times S^3$ and
is completely embedded into the $AdS_7$ part of the background
geometry. We take $(\eta,\a,\b,\th,\g,\d)$ as the worldvolume
coordinates of M5-brane and assume that $\rho$ be only the
function of $\eta$.  For the solution found in \cite{Chen}, $\eta$
and $\rho$ satisfy the following relation:

\begin{equation}
 \sinh\rho=\kappa\sin\eta
\end{equation}
so the induced metric of M5-brane worldvolume is
 \bea
 ds^2&=&\frac{1}{\sin^2\eta}\big(\frac{1+\k^2}{1+\k^2\sin^2\eta}d\eta^2+\cos^2\eta(d\a^2+\sin^2\a
 d\b^2)\big)\nn \\
 & &+\k^2(d\th^2+\sin^2\th d\g^2+\sin^2\th
 \sin^2\g d\d^2).\label{indmetric2}
 \eea

The field strength $H_3$ on the worldvolume is
 \bea
 H_3&=&2a\big(\frac{i}{(1+a^2)\sin^3\eta}\sqrt{\frac{1+\k^2}{1+\k^2\sin^2\eta}}\cos^2\eta\sin\a
 d\eta\w d\a \w d\b\nn\\
& & +\frac{1}{1-a^2}\k^3\sin^2\th\sin\g d\th \w d\g \w d\d\big).
 \eea

The equations of motion require that $\k$ and $a$ should satisfy
\begin{equation}
 {\kappa \over \sqrt{1+\kappa^2}}=-{1-a^2\over 1+a^2}
\end{equation}

\subsection{The computations of the OPE coefficients}
To compute the coupling to the M5-brane of these supergravity
modes, we should compute the variations of the action with respect
to the above fluctuation of the background.

First we notice that after Wick rotation, the non-chiral action for the M5-branes take
the form: \be S=S_{M5}-S_{CS}=T_5\int({1\over 2}\ast {\cal K}+iZ_6).
\ee

We decompose the fluctuation of the background metric into two
parts: \be h_{\um\un}=h^{(1)}_{\um\un}+h^{(2)}_{\um\un},\ee where
 \be h^{(1)}_{\ua\ubb}={1\over 4}g_{\ua\ubb}s,\hs{3ex}
h^{(1)}_{\um\un}=-{1\over 8}g_{\um\un}s.\label{eqh1}  \ee \be
h^{(2)}_{\ua\ubb}=0,\hs{3ex} h^{(2)}_{\um\un}={3\over
8}\d^y_{\um}\d^y_{\un}{1\over y^2}s. \ee Here $s=\sY$.

First we compute the variation of the action with respect to the
first part of the fluctuation of the metric. Let us define \be
\d^{(i)}=h^{(i)}_{\underline{m}\underline{n}}{\d\over \d
g_{\underline{m}\underline{n}}},\hs{3ex} i=1, 2. \ee

From equation (\ref{eqh1}), we get the first part of the
fluctuation of the induced metric as \be
h^{(1)}_{\mu\nu}=-\frac{1}{8}g_{\mu\nu}s,\ee Furthermore, we have
\be \d^{(1)}g^{\mu\nu}=\frac{1}{8}g^{\mu\nu}s. \ee

Since the M5-brane is completely embedded in the $AdS_7$, we have
\begin{eqnarray}
\delta^{(1)}\sqrt{{\rm det}g_{\mu\nu}}&=& {1\over 2}\sqrt{{\rm
det}g_{\mu\nu}}g^{\mu\nu}h^{(1)}_{\mu\nu}\nonumber\\
 &=&-{3\over 8}s\sqrt{{\rm det}g_{\mu\nu}}.
\end{eqnarray}

From
\begin{equation}
H^2=6(H_{\eta\alpha\beta}H^{\eta\alpha\beta}+H_{\theta\gamma\delta}H^{\theta\gamma\delta})
=6(g^{\eta\eta}g^{\a\a}g^{\b\b}H_{\eta\alpha\beta}^2+g^{\th\th}g^{\gamma\gamma}g^{\d\d}H_{\theta\gamma\delta}^2
),\end{equation}
we get
\begin{equation}
\delta^{(1)}H^2=6\cdot 3 \cdot {1\over 8}s
(H_{\eta\alpha\beta}H^{\eta\alpha\beta}+H_{\theta\gamma\delta}H^{\theta\gamma\delta})={3\over
8}sH^2.
\end{equation}

Similarly, by using
\be
H_{mnp}H^{npq}H_{qrs}H^{rsm}= 12
((H_{\eta\alpha\beta}H^{\eta\alpha\beta})^2+(H_{\theta\gamma\delta}H^{\theta\gamma\delta})^2)\ee
we get
\begin{eqnarray}
\delta^{(1)}(H_{mnp}H^{npq}H_{qrs}H^{rsm}) &=&{3\over 4}s
H_{mnp}H^{npq}H_{qrs}H^{rsm}\end{eqnarray}

Since
 \be {\cal
K}=2\sqrt{1+\frac{1}{12}H^2+\frac{1}{288}(H^2)^2-\frac{1}{96}H_{mnp}H^{npq}H_{qrs}H^{rsm}},
\ee we have
\begin{eqnarray}
\delta^{(1)}{\cal K}=0.
\end{eqnarray}
We note that this result is valid for any of the M5-brane solutions completely embedded in
the $AdS_7$ space.
From the above results we get
\begin{eqnarray}
\delta^{(1)}(\sqrt{{\rm det}g_{\mu\nu}}{\cal K})=-{3\over
8}s\sqrt{{\rm det}g_{\mu\nu}}{\cal K}.
\end{eqnarray}
Then
\be \d^{(1)}S_{M5}=-{T_5 \over 2}\int {3\over 8}s\sqrt{{\rm
det}g_{\mu\nu}}{\cal K}d\eta d\a d\b d\th d\gamma d\d. \label{d1m5s}\ee

Now we turn to compute the variation of the action with respect to
the second part of the fluctuation of the background metric.

In Pioncar\`e coordinate, we have
\begin{equation}
h^{(2)}_{\um\un}={3\over 8}\delta^y_{\um} \delta^y_{\un} {1\over
y^2}s.
\end{equation}

In the new coordinate system, we have,
\begin{equation}
h^{(2)}_{\underline{\tilde\mu}\underline{\tilde\nu}}={3\over 8}{\p
y\over \p X^{\underline{\tilde\mu}}} {\p y\over \p
X^{\underline{\tilde\mu}}}{1\over y^2}s,
\end{equation}
which give us
\begin{equation}
h^{(2)}_{\theta\theta}=
h^{(2)}_{\underline{\theta}\underline{\theta}}={3\over
8}s\left({\sinh\rho\sin\theta\over\cosh\rho-\sinh\rho\cos\theta}\right)^2,
\end{equation}
and
\begin{eqnarray}
h^{(2)}_{\eta\eta} &=&{3\over
8}s\left({\cos\eta\over\sin\eta}-{\kappa\cos\eta\over
\cosh\rho}{\sinh\rho-\cosh\rho\cos\theta \over
\cosh\rho-\sinh\rho\cos\theta}\right)^2.
\end{eqnarray}

Similar to the previous computations, we have
\begin{eqnarray}
\delta^{(2)}\sqrt{{\rm det}g_{\mu\nu}}&=& {1\over 2}\sqrt{{\rm
det}g_{\mu\nu}}
g^{\mu\nu}h^{(2)}_{\mu\nu}\nonumber\\
&=& {1\over 2}\sqrt{{\rm det}g_{\mu\nu}}(
g^{\eta\eta}h^{(2)}_{\eta\eta}+
g^{\theta\theta}h^{(2)}_{\theta\theta}),\label{deldets2}
\end{eqnarray}
\begin{eqnarray}
\delta^{(2)}H^2&=&6(\delta^{(2)}
g^{\eta\eta}H_{\eta\alpha\beta}H_\eta^{\,\,\alpha\beta}
+\delta^{(2)}
g^{\theta\theta}H_{\theta\gamma\delta}H_\theta^{\,\,\gamma\delta})\nonumber\\
&=&-6(g^{\eta\eta}h^{(2)}_{\eta\eta}H_{\eta\alpha\beta}H^{\eta\alpha\beta}+
g^{\theta\theta}h^{(2)}_{\theta\theta}H_{\theta\gamma\delta}H^{\theta\gamma\delta}),
\end{eqnarray}
and
\begin{eqnarray}
\delta^{(2)}(H_{mnp}H^{npq}H_{qrs}H^{rsm})&=&12(2(H_{\eta\alpha\beta}H^{\eta\alpha\beta})^2(-
g^{\eta\eta}h^{(2)}_{\eta\eta})\nonumber \\
& &+2(H_{\theta\gamma\delta}H^{\theta\gamma\delta})^2(-
g^{\theta\theta}h^{(2)}_{\theta\theta})).
\end{eqnarray}

Taking all these into account, we get
\begin{eqnarray}
\delta^{(2)}S_{M5}&=&{T_{5}\over 2}\int \sqrt{{\rm det}g_{\mu\nu}}
\left\{h^{(2)}_{\eta\eta} g^{\eta\eta}({{\cal K}\over 2}-{2\over
{\cal K}}H^{\eta\alpha\beta}H_{\eta\alpha\beta}({1\over 2}+{1\over
24}H^2-{1\over
4}H^{\eta\alpha\beta}H_{\eta\alpha\beta}))\right.\nonumber\\
& &+\left. (\eta, \alpha, \beta\to \theta, \gamma, \delta)\right\}
\mathrm{d}\eta\mathrm{d}\alpha\mathrm{d}\beta\mathrm{d}\theta\mathrm{d}\gamma\mathrm{d}\delta\nonumber\\
&=&-{T_{5}\over
2}\int\mathrm{d}\eta\mathrm{d}\alpha\mathrm{d}\beta\mathrm{d}\theta\mathrm{d}\gamma\mathrm{d}\delta
{3\over 8}s{\cos^2\eta+\sinh^2\rho\sin^2\theta \over
(\cosh\rho-\sinh\rho\cos\theta)^2}\nn\\& & \times
{\kappa^2\cos^2\eta\sin\alpha\sin^2\theta\sin\gamma\over
\sin^3\eta\cosh\rho},\label{d2m5s}
\end{eqnarray}
where $\kappa={1-a^2\over 2|a|}$ and the explicit value of $H_3$
have been used.

Now we begin to discuss the contributions from the fluctuation of
the four-form flux. Recall that $\d H_3=-\d\underline{C}_3$, since
$\d C_3$ only have components in $S^4$, so $\delta H_3=0$. Then
the contributions  only come from the Chern-Simions part of the
action. Since $\delta (\underline{C}_3\w H_3)=0$, the only
contribution is from $\delta \underline{C}_6$.

The computations of $\delta C_6$ is put in the appendix, the
result is \be \d C_6=-{3\over 8 }C_6 \sum_I s^IY^I.\ee Therefore
\be \d S_{CS}=-iT_5\int\d \underline{C}_6=-iT_5 \int
\underline{C}_6(-{3\over 8}s).\ee Using this result and
eq.~(\ref{d1m5s}), we get \be\d^{(1)}S_{M5}-\d S_{CS}=T_5\int
({1\over 2}\star{\cal K}+i\underline{C}_6)(-{3\over 8}s)\ee

The 6-form gauge potential $C_6$ is of the form \bea
 C_6&=&i\frac{\cos^3\eta\sinh^3\rho\sin^2\th\sin\a\sin\g}{\sin^6\eta}d\rho \w
  d\a\w d\b \w d\th \w d\g\w d\d \nn\\
  &  &-i\frac{\cos^2\eta\sinh^2\rho\sin^3\th\sin\a\sin\g}{\sin^5\eta(\cosh\rho-\sinh\rho\cos\th)}d\eta
  \w d\a \w d\b \w d\rho\w d\g\w d\d \nn\\
  &  &+i\frac{\cos^2\eta\sinh^3\rho\sin^2\th\sin\a\sin\g(\sinh\rho-\cos\th\cosh\rho)}{\sin^5\eta(\cosh\rho-\sinh\rho\cos\th)}\nn\\
& &  d\eta
  \w d\a \w d\b \w d\th\w d\g\w d\d.
  \eea
On the M5-brane worldvolume, \bea
\underline{C}_6&=&i{\kappa^3\cos^2\eta\sin^2\theta\sin\alpha\sin\gamma\over
\sin^3\eta\cosh\rho(\cosh\rho-\sinh\rho\cos\theta)}(\kappa\cosh\rho-(1+\kappa^2)\cos\theta\sin\eta)\nn\\
& & d\eta\wedge d\a\wedge d\b \w d\th \w d\gamma\w d\d \eea

After some calculations, we get \bea\d^{(1)}S_{M5}-\d S_{CS}&=&T_5
\int\left(-{3\over
8}s\right){\kappa^2\cos^2\eta\sin^2\theta\sin\alpha\sin\gamma\over
2
\sin^3\eta\cosh\rho}{\cosh\rho+\sinh\rho\cos\theta\over\cosh\rho-\sinh\rho\cos\theta}\nn\\
& &d\eta d\alpha d\beta d\theta d\gamma d\d
 \eea
From this result and eq.~(\ref{d2m5s}), we get \bea \d
S&=&\d^{(1)}S_{M5}+\d^{(2)}S_{M5}-\d S_{CS}=T_5\int {3\over
8}s{\kappa^2\cos^2\eta\sin^2\theta\sin\alpha\sin\gamma\over
2\sin^3\eta\cosh\rho}\nn\\
& &{\sin^2\eta-2\sinh^2\rho\sin^2\theta-2\over
(\cosh\rho-\sinh\rho\cos\theta)^2} d\eta d\a d\b d\theta d\gamma
d\d \eea

Having obtained the variation of the action with respect to the
fluctuation of the background fields, we can compute the
correlation function of the Wilson surface operator in the
symmetric representation with the chiral primary operators.

Now, we write $s^I$ as  $s^I(\vec{x}, y)=\int d^6\vec{x}^\prime
G_\Delta(\vec{x}^\prime; \vec{x}, y)s^I_0(\vec{x}^\prime)$,
\bea{\langle W({\cal S}, L) {\cal O}_\Delta(0) \rangle\over
\langle W({\cal S}, L) \rangle} &\sim& -{1\over {\cal N}^I}{\d
S\over \d s^I_0(\vec{x})}=-{T_5 \over {\cal N}^I}\int {3\over 8}c
{r^\Delta\kappa^{1-\Delta}\sinh^\Delta\rho\cos\eta\sin^2\theta\sin\alpha\sin\gamma
\over
2L^{2\Delta}\sin^3\eta(\cosh\rho-\sinh\rho\cos\theta)^{\Delta+2}
}\nonumber \\
&\times&(\sin^2\eta-2\sinh^2\rho\sin^2\theta-2)Y^I(\tilde\theta)d\rho
d\a d\b d\th d\gamma d\d.\eea

By using \be \int_0^\pi d\a\sin\a \int_0^{2\pi}d\b= \int_0^\pi
d\gamma\sin\gamma \int_0^{2\pi}d\d=4\pi \ee and \be
\sin\eta=\kappa^{-1}\sinh\rho,
\hs{3ex}\cos\eta={\sqrt{\kappa^2-\sinh^2\rho}\over \kappa}, \ee we
get \bea{\langle W({\cal S}, L) {\cal O}_\Delta(0) \rangle\over
\langle W({\cal S}, L) \rangle} &\sim& -{c\over {\cal N}^I}\cdot
(4\pi)^2 {3\over 8}{r^\Delta \over L^{2\Delta}}
T_5\int_0^{\sinh^{-1}\kappa}{1\over
2}\sqrt{\kappa^2-\sinh^2\rho}\kappa^{3-\Delta}\sinh^{\Delta-3}
\rho d\rho\nn \\ & & \int_0^\pi d\theta
{\sin^2\theta(-2+\sinh^2\rho(\kappa^{-2}-2\sin^2\theta))\over
(\cosh\rho-\sinh\rho\cos\theta)^{2+\Delta} }Y^I(\tilde\theta).\eea

So the OPE coefficient is \bea c_{S,
\Delta}&=&2^{3k/2+4}(k+{1\over
2})\pi^{-5/4}N^{1/2}\sqrt{{(2k-1)\Gamma(k+1/2)\over
\Gamma(k)}}\nn\\
&
&\int_0^{\sinh^{-1}\kappa}\sqrt{\kappa^2-\sinh^2\rho}\kappa^{3-\Delta}\sinh^{\Delta-3}
\rho d\rho\nn \\ & & \int_0^\pi d\theta
{\sin^2\theta(-2+\sinh^2\rho(\kappa^{-2}-2\sin^2\theta))\over
(\cosh\rho-\sinh\rho\cos\theta)^{2+\Delta} }.\eea We can perform
the integral over $\theta$ and get: \bea c_{S,
\Delta}&=&2^{3k/2+2}(k+{1\over
2})\pi^{-1/4}N^{1/2}\sqrt{{(2k-1)\Gamma(k+1/2)\over
\Gamma(k)}}\nn\\
&
&\int_0^{\sinh^{-1}\kappa}d\rho\sqrt{\kappa^2-\sinh^2\rho}\kappa^{3-\Delta}\sinh^{\Delta-3}
\rho \nn \\ & & [2(\kappa^{-2}\sinh^2\rho-2)\exp[-(2+\Delta)\rho]\ _2F_1(3/2, 2+\Delta, 3, 1-e^{-2\rho})\nn\\
& & -3\sinh^2\rho \exp[-(2+\Delta)\rho]\ _2F_1(5/2, 2+\Delta, 5,
1-e^{-2\rho})].\eea

It would be interesting to compare our results with the OPE
coefficients of Wilson surface operators in the fundamental
representation computed using the membranes \cite{Corrado99}. To
do this, we should take the limit of $\kappa\to 0$ because in this
limit the $S^3$ part of the worldvolume shrink.

In this limit, we can do the integral by substitution: we define
$t$ by using \be \rho=(\sinh^{-1}\kappa) t,\hs{3ex} 0\le t\le 1
\ee then as $\kappa\to 0$, \be\sinh^{-1}\kappa\sim \kappa,\hs{3ex}
\rho\sim\kappa t,\hs{3ex} \cosh\rho\sim 1,
\hs{3ex}\sinh\rho\sim\kappa t, \hs{3ex}d\rho\sim\kappa dt,
 \ee
then \bea c_{S, \Delta}&=&-{1\over {\cal N}^I}\cdot 3\pi^2 c
T_5\kappa^2\int_0^1dt t^{\Delta-3}
(t^2-2)\sqrt{1-t^2} \int_0^\pi d\theta\sin^2\theta\nn\\
&=&-{3\pi^3\over 2}{c\over {\cal N}^I}{a^\Delta\over
L^{2\Delta}}T_5\kappa^2\int_0^1dt\sqrt{1-t^2}(t^{\Delta-1}-2t^{\Delta-3}).\eea

Using
\bea
\int_0^1dt\sqrt{1-t^2}(t^{\Delta-1}-2t^{\Delta-3})&=&\frac{\sqrt{\pi}}{4}(\frac{\Gamma({\Delta\over
2})}{\Gamma({\Delta+3\over 2})}-2\frac{\Gamma({\Delta-2\over
2})}{\Gamma({\Delta+1\over 2})})\nn\\
&=&-{\sqrt{\pi}\over 4}{\Delta+4\over
\Delta+1}\frac{\Gamma({\Delta-2\over 2})}{\Gamma({\Delta+1\over
2})}, \eea we get \bea c_{S, \Delta}&=& T_5 {c\over {\cal N}^I}
{3\pi^{7/2}\over 8}  \kappa^2 {2k+4\over
2k+1}\frac{\Gamma(k-1)}{\Gamma(k+{1\over 2})}\nn\\
&=&-2^{(3k+3)/2}N^{1/2}\pi^{1/4}{k+2 \over
k-1}\sqrt{{\Gamma(k)\over \Gamma(k-1/2)}}\kappa^2,\eea in the
$\kappa\to 0$ limit. Now we express this result in terms of
$Q_M$,
 the magnetic charge of the string soliton solution\cite{Chen}.
For this solution, we have $\kappa^2=Q_M/(8\pi N)$, So \be c_{S,
\Delta}=-Q_M2^{(3k-3)/2}\pi^{-3/4}{k+2 \over
k-1}\sqrt{{\Gamma(k)\over N\Gamma(k-1/2)}}.\ee

We can see that in this limit the OPE coefficients is proportional
to $Q_M$. Comparing with the results eq.~(\ref{eqcf}) obtained
from membrane, we find that the $k$-dependence of the OPE
coefficient is different although the $N$-dependence is the same.

\section{OPE of the Wilson surface in the antisymmetric representation}

In this section we compute the OPE of the Wilson surface in the
antisymmetric representation. As mentioned in the introduction, in
this case, the worldvolume of the M5-brane still has topology
$AdS_3\times S^3$, where the $S^3$ part (we sometimes call it
$\tilde S^3$) is in $S^4$ instead of $AdS_7$. Some part of the
calculations are similar to the previous section, while some new
issues will appear here.\footnote{In this section, we set the
vector $\tilde\theta^I$ mentioned in section 2 to be $(1, 0, 0,
0)$ by a $SO_R(5)$ rotation. Then the corresponding angular
coordination $\z_1$ equals to zero. }

\subsection{Review of the M5-brane solution}
As in section 4, we first review the M5-brane solution
corresponding to the spherical Wilson surface operator in
antisymmetric representation.

We begin from the Euclidean $AdS_7$ whose metric has form
(\ref{metric7}). We further consider the transformation
\begin{equation}
y=r\cos\d,\hs{3ex} r_1=r\sin\d.
\end{equation}
The coordinates of the $AdS_3$ part of the M5-brane worldvolume can
be chosen as $\d, \a, \b$. Then the $AdS_3$ part of the induced
metric of the worldvolume is
 \be\label{indmetric4}
 ds^2_{\mbox{\small ind, $AdS_3$}}=\frac{1}{\cos^2\d}(d\d^2+\sin^2\d(d\a^2+\sin^2\a
 d\b^2)).
 \ee
The coordinates of the $\tilde S^3$ part can be chose to be $\z_2,
\z_3, \z_4$ and we let $\z_1$ to be fixed at a constant $\z^0$.
Then the induced metric of this part is \be ds^2_{\mbox{\small
ind, $\tilde S^3$}}={1\over
4}\sin^2\z^0(d\z_2^2+\sin^2\z_2d\z_3^2+\sin^2\z_2\sin^2\z_3d\z_4^2)
\ee The field strength $H_3$ on the worldvolume is \bea
H_3&=&2a\left(i{1\over 1+a^2}{\sin^2\d\sin\a\over \cos^3\d} d\d\w
d\a \w d\b\right.\nn\\& &\left.+{1\over 1-a^2}{\sin^3\z^0\over
8}\sin^2\z_2\sin\z_3 d\z_2\w d\z_3 \w d\z_4
 \right).\eea
The equations of motion require $a$ and $\z^0$ should satisfy \be
a={\pm 1+\sin\z^0\over \cos\z^0}. \ee

\subsection{The computations of the OPE coefficients}

After reviewing the M5-brane solution, we now compute the OPE
coefficients of the Wilson surface operators using $AdS_7/CFT_6$
correspondence. As the computation for Wilson surfaces in the
symmetric representation, we should compute the variation of the
M5-brane action with respect to the fluctuation of the background
fields reviewed in section 2.

For the variation with respect to the first part of the
fluctuation of the metric, we have:
 \be \d^{(1)}\sqrt{{\rm det}
g_{mn}}={1\over 2}\sqrt{{\rm det}
g_{mn}}(g^{\a\b}h^{(1)}_{\a\b}+g^{\mu\nu}h^{(1)}_{\mu\nu})={1\over
2}\sqrt{{\rm det} g_{mn}}(-{3\over 8}\sY+{3\over 4}\sY).\ee

From
\begin{equation}
H^2=6(H_{\d\alpha\beta}H^{\d\alpha\beta}+H_{234}H^{234}),
\end{equation}
and \be H_{mnp}H^{npq}H_{qrs}H^{rsm}
=12\cdot((H_{\delta\alpha\beta}H^{\delta\alpha\beta})^2+(H_{234}H^{234})^2),\ee
we get

\begin{equation}
\delta^{(1)}H^2=6({3\over 8}\sY
H_{\d\alpha\beta}H^{\d\alpha\beta}-{3\over 4 }\sY H_{234}H^{234}),
\end{equation}
and \be \delta^{(1)}(H_{mnp}H^{npq}H_{qrs}H^{rsm})
=12\cdot\left(2\cdot {3\over
8}\sY\cdot(H_{\delta\alpha\beta}H^{\delta\alpha\beta})^2-2\cdot{3\over
4}\sY(H_{234}H^{234})^2\right).\ee So
\begin{eqnarray}
&&\delta^{(1)}({1\over 12}H^2+{1\over 288}(H^2)^2-{1\over
96}H_{mnp}H^{npq}H_{qrs}H^{rsm})\nonumber\\
&=&{3\over 8}\sY H_{\d\a\b}H^{\d\a\b}({1\over 2 }+{1\over
24}H^2-{1\over 4}H_{\d\alpha\beta}H^{\d\alpha\beta})\nn\\
& &-{3\over 4}\sY H_{234}H^{234}({1\over 2}+{1\over 24}H^2-{1\over
4}H_{234}H^{234}).\eea From
 \be {\cal
K}=2\sqrt{1+\frac{1}{12}H^2+\frac{1}{288}(H^2)^2-\frac{1}{96}H_{mnp}H^{npq}H_{qrs}H^{rsm}},\ee
we have \be \delta^{(1)}({\cal K}) ={2 \over {\cal K}
}\delta^{(1)}(\frac{1}{12}H^2+\frac{1}{288}(H^2)^2-\frac{1}{96}H_{mnp}H^{npq}H_{qrs}H^{rsm}).\ee
Using this, we get
\begin{eqnarray}
\delta^{(1)}(\sqrt{{\rm det}g_{mn}}{\cal K})&=&\sqrt{{\rm
det}g_{mn}}[-{3\over 8}\sY({{\cal K}\over 2}-{2\over {\cal
K}}H_{\d\a\b}H^{\d\a\b}({1\over 2 }+{1\over 24}H^2-{1\over
4}H_{\d\alpha\beta}H^{\d\alpha\beta}))\nn \\& &+{3\over
4}\sY({{\cal K}\over 2}-{2\over {\cal K}}H_{234}H^{234}({1\over 2
}+{1\over
24}H^2-{1\over 4}H_{234}H^{234}))]\nn\\
&=& \sqrt{{\rm det}g_{mn}}(-{3\over 8}\sY (-{1+a^2\over
1-a^2})+{3\over 4}\sY(-{1-a^2\over 1+a^2}))\nn\\
&=& {3\over 8}\sY \sqrt{{\rm det}g_{mn}}{-1+6a^2-a^4\over
1-a^4}.\label{d1sga}
\end{eqnarray}

Now $y=r\cos\d$, so \be h^{(2)}_{\d\d}={3\over 8}s \left({\p
y\over \p \d}\right)^2{1\over y^2}={3\over 8}s{\sin^2\d\over
\cos^2\d}.\ee Similar to the computations for the Wilson surface
operators in symmetric representation, we have  \bea
\delta^{(2)}(\sqrt{{\rm det}g_{mn}}{\cal K})&=&\sqrt{{\rm
det}g_{mn}}g^{\d\d}h^{(2)}_{\d\d}({{\cal K}\over 2}-{2\over {\cal
K}}H_{\d\a\b}H^{\d\a\b}({1\over 2 }+{1\over 24}H^2-{1\over
4}H_{\d\alpha\beta}H^{\d\alpha\beta}))\nn\\
&=& \sqrt{{\rm det}g_{mn}} {3\over 8}s\sin^2\d {a^2+1\over
a^2-1}\label{d2sga}\eea

So from eqs.~(\ref{d1sga}) and (\ref{d2sga}), we get the
contribution from the fluctuation of the metric:
 \be\d_g(\sqrt{{\rm
det}g_{mn}}{\cal K})=\sqrt{{\rm det}g_{mn}}{3\over
8}\sY({-1+6a^2-a^4\over 1-a^4} +\sin^2\d {a^2+1\over a^2-1}).\ee

Now we turn to the contribution from the background flux. Unlike
the symmetric case, the pullback of $\d C_3$ on the worldvolume is
nonzero, then we will get a contribution from $\d H$. In fact,
from
\begin{equation}
\delta C_{\ua \ubb \ug}=\sum_{I}{3\over 16k }\epsilon_{\ua \ubb
\ug \ud}s^I \nabla^{\ud} Y^I,
\end{equation}
we get
 \be \d C_{\underline{234}}=-{3\over
16k}\sin^3\z_1\sin^2\z_2\sin\z_3\sum_I s^I\p_{\z_1}Y^I, \ee then
\be \d \underline{C}_{234}=-{3\over
16k}\sin^3\z^0\sin^2\z_2\sin\z_3\sum_I s^I\p_{\z^0}Y^I. \ee

From  \be\d H_3=-\d\underline{C}_3, \ee we get \be \d
H_{234}={3\over 16k}\sin^3\z^0\sin^2\z_2\sin\z_3\sum_I
s^I\p_{\z^0}Y^I. \ee

Recall that \be H_{234}={a\over
4(1-a^2)}\sin^3\z^0\sin^2\z_2\sin\z_3, \ee we have\footnote{Here
$H_{234}$ denotes $H_{\z_2\z_3\z_4}$.} \be \d
H_{234}={3(1-a^2)\over 4ak}H_{234}\sum_I s^I\p_{\z^0}Y^I. \ee

From this we can easily get \be\d_H
H^2=6g^{22}g^{33}g^{44}H_{234}\cdot 2\d H_{234}=12
H_{234}H^{234}{\d H_{234}\over H_{234}}, \ee and \be
\delta_H(H_{mnp}H^{npq}H_{qrs}H^{rsm})=12\cdot 4
(H_{234}H^{234})^2{\d H_{234}\over H_{234}}. \ee Then \bea
&&\delta_H({1\over 12}H^2+{1\over 288}(H^2)^2-{1\over
96}H_{mnp}H^{npq}H_{qrs}H^{rsm})\nonumber\\
&=& {3a(1+a^4)\over k(1-a^2)(1+a^2)^2}\sum_I s^I\p_{\z^0}Y^I.\eea
Putting all these together, we have, \be\d_H(\sqrt{{\rm
det}g_{mn}}{\cal K})=\sqrt{{\rm det}g_{mn}}\d_H\left(\frac{{\cal
K}^2}{4} \right)\frac{2}{\cal K}=-\sqrt{{\rm det}g_{mn}}{3a\over
k(a^2+1)}\sum_I s^I\p_{\z^0}Y^I. \ee

Finally let us compute the variation of the Chern-Simions term.
Recall that \be Z_6=\underline{C}_6-{1\over 2}\underline{C}_3\w
H_3.\ee In this case, $\d \underline{C}_6=0$, so
\bea \d Z_6&=&-{1\over 2}\d\underline{C}_3\w H_3\nn\\
&=& -{3ia\over
16k}{\sin^3\z^0\sin^2\z_2\sin\z_3\sin\alpha\sin^2\delta\over
\cos^3\delta(1+a^2)}\sum_Is^I\p_{\z^0}Y^I\nn\\
& &d\d\w d\a\w d\b\w d\z_2\w d\z_3\w d\z_4.\eea \

The total action of M5-brane is \be
S=S_{M5}-S_{CS}=T_5\int(\star{1\over 2}{\cal K}+iZ_6),\ee so we get
\bea \d S&=&T_5\int(({1\over 2}\delta_g(\sqrt{{\rm
det}g_{mn}}{\cal K})+{1\over 2}\sqrt{{\rm
det}g_{mn}}\delta_H({\cal K}))d\d\w d\a \w d\b \w d\z_2 \w
d\z_3\w d\z_4\nn\\
& &-\d Z_6)\nn\\
&=& {3T_5\over 8}
\int{\sin^3\z^0\sin^2\z_2\sin\z_3\sin\alpha\sin^2\delta\over
16\cos^3\delta}({-1+6a^2-a^4\over 1-a^4} +\sin^2\d
{a^2+1\over a^2-1})\nn\\
& & \times\sum_Is^IY^Id\d d\a d\b d\z_2 d\z_3 d\z_4.\nn\\ &=&
{3T_5\over 8}
\int{\sin^3\z^0\sin^2\z_2\sin\z_3\sin\alpha\sin^2\delta\over
16\cos^3\delta}({-1+6a^2-a^4\over 1-a^4} +\sin^2\d
{a^2+1\over a^2-1})\nn\\
& & \times\sum_ks^kY^kd\d d\a d\b d\z_2 d\z_3 d\z_4. \eea Here we
have performed the integration over the 3-sphere\footnote{Here
$Y^{k, 0}$ is the abbreviation of $Y^{(k, 0, 0, 0)}$. For
discussions on spherical harmonics, see, for example,
\cite{Higuchi:1986wu}.}: \be\int\sin^2\z_2\sin\z_3\sum_I
s^IY^Id\z_2d\z_3d\z_4=\sum_k s^k Y^{k, 0}(\z^0).\ee

As before, using \be G_\Delta(\vec{x}^\prime; \vec{x}, y)\simeq
c{y^\Delta \over L^{2\Delta}},\hs{3ex} y=r\cos\delta,\ee and
\be\int_0^\pi\sin\a d\a \int_0^{2\pi}d\b=4\pi, \ee we get
\bea{\langle W({\cal S}, L) {\cal O}_\Delta(0) \rangle\over
\langle W({\cal S}, L) \rangle} &\sim&-\frac{3\pi T_5}{32}{c \over
{\cal N}^I}\sum_\Delta{r^\Delta\over
L^{2\Delta}}\int_0^{\pi/2} {\sin^3\z^0\sin^2\delta\cos^{\Delta-3}}\d\nn\\
& & ({a^4-6a^2+1\over
a^4-1}+{\sin^2\delta{a^2+1\over a^2-1}})Y^{k, 0}(\z^0).
 \eea

Now we perform the integration over $\d$: \be\int_0^{\pi/2}
\sin^2\d\cos^{\Delta-3}\d d\d={\sqrt{\pi}\over
4}{\Gamma({\Delta\over 2}-1)\over \Gamma({\Delta+1\over 2})}, \ee

\be\int_0^{\pi/2} \sin^4\d\cos^{\Delta-3}\d d\d={\sqrt{\pi}\over
4}{3\over \Delta+1}{\Gamma({\Delta\over 2}-1)\over
\Gamma({\Delta+1\over 2})}, \ee and have \bea{\langle W({\cal S},
L) {\cal O}_\Delta(0) \rangle\over \langle W({\cal S}, L) \rangle}
&\sim& \mp \frac{3\pi^{3/2} T_5}{128}{c\over {\cal
N}^I}\sum_\Delta{r^\Delta\over
L^{2\Delta}}\sin^3\z^0(-{\cos2\z^0\over \sin\z^0}\nn\\
& &+{{1\over \sin\z^0}}{3\over 2k+1})Y^{k,
0}(\z^0){\Gamma(k-1)\over \Gamma(k+1/2)}, \eea after putting the
explicit value of $a$.

The harmonic function can be written as \be Y^{k, 0}(\z^0)={\cal
N}_k C^{(3/2)}_k(x), \ee where $x=\cos\z^0$, $C^{(3/2)}_k(x)$ are
Gegenbauer polynomials and \be {\cal N}_k=\left[{\pi^{1/2}
k!(2k+3)\over 2^{3k+7}(k+1)(k+2)\Gamma(k+5/2)}\right]^{1/2}, \ee
is obtained from the normalization of $Y^{k, 0}.$

 Therefore \bea{\langle
W({\cal S}, L) {\cal O}_\Delta(0) \rangle\over \langle W({\cal S},
L) \rangle} &\sim& \mp  \frac{3\pi^{3/2} T_5}{128}{c\over {\cal
N}^I}\sum_\Delta{r^\Delta\over
L^{2\Delta}}Y^{k, 0 }(0)\sin^2\z^0((-{\cos2\z^0}\nn\\
& &+{3\over 2k+1}){{\cal N}_k \over Y^{k, 0}(0)}
C^{(3/2)}_{k}(\cos\z^0) ){\Gamma(k-1)\over \Gamma(k+1/2)}. \eea

From \be Y^{k, 0}(0)={\cal N}_k  C^{(3/2)}_k(1),\ee we have
\bea{\langle W({\cal S}, L) {\cal O}_\Delta(0) \rangle\over
\langle W({\cal S}, L) \rangle} &\sim&  \mp \frac{3\pi^{3/2}
T_5}{128}{c\over {\cal N}^I}\sum_\Delta{r^\Delta\over
L^{2\Delta}}Y^{k, 0 }(0) {1 \over C^{(3/2)}_k(1)}\sin^2\z^0\nn\\&
& \times(-{\cos2\z^0}+{3\over 2k+1})
C^{(3/2)}_k(\cos\z^0){\Gamma(k-1)\over \Gamma(k+1/2)}. \eea

So the OPE coefficients is \bea c_{A, \Delta} &\sim& \pm
 \frac{2^{(3k-5)/2}N^{1/2}}{\pi^{7/4}}{C^{(3/2)}_k(\cos\z^0) \over
C^{(3/2)}_k(1)}\sin^2\z^0\nn\\& & \times(-{\cos2\z^0}+{3\over
2k+1}) {k+1/2 \over k-1} \sqrt{\Gamma(k)\over \Gamma(k-1/2)}. \eea

To compare with the membrane results, we take the $\z^0\to 0$ limit in which
the $\tilde S^3$ will shrink.
In this limit $x\to 1$ and the OPE cooeficient is equal to \be \mp
 \frac{2^{(3k-5)/2}N^{1/2}}{\pi^{7/4}}(\z^0)^2\sqrt{\Gamma(k)\over \Gamma(k-1/2)}\ee

The magnetic charge of string soliton solution is \be
 Q_M=\frac{1}{{\mbox{Vol}(\tilde S^3)}}\int_{\tilde S^3}  H=
-\frac{\sin^2\z^0\cos\z^0}{8l^3_p}, \ee in the small $\z^0$ limit,
we have \be
 Q_M=
-\frac{(\z^0)^2}{8l^3_p}=-\pi N(\z^0)^2. \ee Then the OPE
coefficients can be written as \be \pm
 \frac{2^{(3k-5)/2}}{\pi^{11/4}}Q_M\sqrt{\Gamma(k)\over N\Gamma(k-1/2)}.\ee
We can see that in this limit the OPE coefficients is proportional
to $Q_M$ and
 the $k$-dependence and $N$-dependence of the
coefficients in this limit is the same as the one in
eq.~(\ref{eqcf}) computed using M2-brane.

\section{Conclusion and discussions}

In this paper we studied the OPE of spherical half-BPS Wilson
surface operators using their M5-brane description. We computed
the OPE coefficients by studying the coupling to the M5-branes of
the supergravity modes. In this process, we first make clear that
the variation of the embedding and the 2-form gauge potential can
be set to zero. Then we calculated the response of the non-chiral
action of M5-brane to the bulk supergravity fields. Moreover, we
had to investigate carefully the response of the Chern-Simons term
in the M5-brane action  to the bulk gauge potential. In the
symmetric case, the three form field strength has no fluctuation
and only the fluctuation of the dual 6-form gauge potential gives
the contribution. On the contrary, in the antisymmetric case,
$\delta H_3$ is non-zero while $\d \underline{C}_6=0$.

We also consider the membrane limit of our results. In this limit
the $S^3$ part of the M5-brane worldvolume shrink. We find that
the OPE coefficients is proportional to $Q_M$ which characterizes
the rank of the representation. This is reminiscent of the results
for the expectation values of these Wilson surfaces in
\cite{Chen}. There it is found that the expectation values is
proportional to $Q_M$ even before we take the membrane limit. We
compare our result in this membrane limit with the results
obtained from the membrane method \cite{Corrado99}. We find that
the $N$ dependence are the same. We also find that for the Wilson
surface in symmetric representation, the dependence on the
dimension of the local operator is different, while in the
antisymmetric case, the dependence is the same. This may be
related to the nontrivial dynamics of the branes in M-theory. We
hope we can come back to this point in the future.

Another subtle issue is the choice of the M5-brane action.  Among
the different proposals for the M5-brane action, we chose the
non-chiral action since there are no auxiliary fields in this
action. Although different choice of action gives equivalent
equations of motion \cite{Sorokin97, Sezgin99}, this does not
guarantee that these actions give the same quantum dynamics. It
will be interesting to compute the OPE coefficients using other
actions of the M5-brane, such as the PST action and compare the
results obtained from different action. In \cite{Chen}, the issue
on choosing action also appear and make the discussions for the
boundary terms quite subtle.

At this stage, quite little is known in the field theory side.
Some field theory studies could be found in \cite{Gustavsson}: the
conformal anomaly of abelian Wilson surface operator was
calculated in $A_1$ field theory. It is very hard to consider the
nonabelin Wilson surface in field theory. It would be interesting
to study the OPE of Wilson surface operators from field theory
calculation and compare the results with the ones obtained from
M2-brane or M5-brane.

If we compactify the six-dimension (0, 2) SCFT on $T^2$ with
supersymemtric boundary conditions, we will obtain ${\cal N}=4$
supersymmetric Yang-Mills theory. If the Wilson surface winds
various 1-cycles of the $T^2$, it will give Wilson loop, 't Hooft
loop or Wilson-'t Hooft loop \cite{Henningson:2000nu,
Arvidsson:2006rv}. It is interesting to see if one can study this
relation in the framework $AdS/CFT$ correspondence. The relation
between Wilson surface in six-dimensional SCFT and the surface
operator in four-dimensional SYM is also a quite interesting
subject \cite{Gukov:2006jk, Constable:2002xt, Gomis:2007fi,
Buchbinder:2007ar}.

Another interesting subject about the Wilson surfaces in higher
dimensional representation is to compute the correlation function
of two Wilson surfaces, one in the fundamental representation and the
other one in higher dimensional representation\cite{ChenLiuWu2}.

\section*{Acknowledgments}
The work was partially supported by NSFC Grant No. 10535060,
10775002 and NKBRPC (No. 2006CB805905). JW would like to thank the
Galileo Galilei Institute for Theoretical Physics, Prof. Chuan-Jie
Zhu, Institute of Theoretical Physics, Chinese Academy of Science
and KITPC, CAS for their hospitality. JW also thanks Jarah Evslin
and Giulio Bonelli for helpful discussions. The work of JW is
supported in part by the European Community's Human Potential
Programme under contract MRTN- CT-2004-005104 `Constituents,
fundamental forces and symmetries of the universe' as a postdoc of
the node of Padova at the initial stage of the project.

\section{Appendix: The variation of $C_6$ due to the SUGRA modes}

In this appendix we compute  $\delta C_6$ due to the supergravity
modes $s^I$. Notice that for the purpose of this paper, we can use
the approximation eq.~(\ref{eqy}) freely here.

From
\begin{equation}
\delta C_{\ua \ubb \ug}=\sum_{I}{3\over 16k }\epsilon_{\ua \ubb \ug
\ud}s^I \nabla^{\ud} Y^I,
\end{equation}
we get \be\delta H_{\um\ua\ubb\ug}={3\over 16k}\sum_I
\epsilon_{\ua\ubb\ug\ud}\nabla^{\ud}Y^I\nabla_{\um}s^I,\ee and
\be\delta H_{\ua\ubb\ug\ud}=-{3\over 16k}\sum_I
\epsilon_{\ua\ubb\ug\ud}\nabla_{\underline{\epsilon}}
\nabla^{\underline{\epsilon}}Y^Is^I.\ee We notice that \be
\nabla_{\underline{\epsilon}}
\nabla^{\underline{\epsilon}}Y^I=-4k(k+3)Y^I,\ee so we get
\be\delta H_{\ua\ubb\ug\ud}={3(k+3)\over 4}\sum_I
\epsilon_{\ua\ubb\ug\ud}Y^Is^I,\ee considering \be
H_{\ua\ubb\ug\ud}=6\epsilon_{\ua\ubb\ug\ud},\ee \be\delta
H_{\ua\ubb\ug\ud}={(k+3)\over 8}\sum_I H_{\ua\ubb\ug\ud}Y^Is^I.\ee

Since $H_7$ is the Hodge dual of $H_4$: \be
(H_7)_{\underline{m}_1\cdots \underline{m}_7}={\sqrt{g}\over
4!}\epsilon^{\underline{n}_1\cdots
\underline{n}_4}_{\,\,\,\,\,\,\,\,\,\,\,\,\,\,\,\,\underline{m}_1\cdots
\underline{m}_7} H_{\underline{n}_1\cdots \underline{n}_4},\ee we
get \bea (\delta H_7)_{\um_1\cdots\um_6\ua}&=&{\sqrt{g}\over
4!}4\epsilon^{\ua_1\ua_2\ua_3\um}_{\,\,\,\,\,\,\,\,\,\,\,\,\,\,\,\,\,\,\,\,\,\,\um_1\cdots
\um_6\ua}\, \delta H_{\ua_1\ua_2\ua_3\um} \nn\\
&=&{3\over 16k}\epsilon_{\um\um_1\cdots\um_6}\sum_I
\nabla_{\ua}Y^I\nabla^{\um}s^I \nn\\
&\simeq&{3\over 8}y\epsilon_{y\um_1\cdots\um_6}\sum_I
s^I\nabla_{\ua}Y^I,\eea and \bea (\d H_7)_{\um_1\cdots
\um_7}&=&{\sqrt{g}\over 4!}\epsilon^{\ua_1\cdots
\ua_4}_{\,\,\,\,\,\,\,\,\,\,\,\,\,\,\,\,\um_1\cdots \um_7} \delta
H_{\ua_1\cdots \ua_4}+{\d\sqrt{g}\over 4!}\epsilon^{\ua_1\cdots
\ua_4}_{\,\,\,\,\,\,\,\,\,\,\,\,\,\,\,\,\um_1\cdots \um_7}
H_{\ua_1\cdots \ua_4}\nn\\
& &+\left({\sqrt{g}\over 4!}\d
g^{\ua_1\ubb_1}g^{\ua_2\ubb_2}g^{\ua_3\ubb_3}g^{\ua_4\ubb_4}\varepsilon_{\ubb_1\cdots
\ubb_4\um_1\cdots \um_7} \delta H_{\ua_1\cdots \ua_4}\right.\nn\\&
&+\left.\mbox{three other terms from $\d g^{\ua_i\ubb_i}, i=2, 3,
4$}\right).\eea By using \bea \delta \sqrt{g}&=&{1\over
2}\sqrt{g}g^{\underline{m}\underline{n}}\d
g_{\underline{m}\underline{n}}={1\over
2}\sqrt{g}(g^{\ua\ubb}\d g_{\ua\ubb}+g^{\um\un}\d g_{\um\un}) \nn\\
&=&{1\over 4}\sqrt{g}s, \eea and \be \d g^{\ua\ubb}=-{1\over
4}g^{\ua\ubb}s, \ee we get \bea (\d H_7)_{\um_1\cdots\um_7} &=& {k-3
\over 8}(H_7)_{\um_1\cdots\um_7}\sum_I s^IY^I\eea From eq.~(\ref{flux}) and $\d (C_3\w
H_4)=0$, we get $d\d C_6=\d H_7$. We can choose \be \d C_6=-{3\over 8
}C_6 \sum_I s^IY^I,\ee which lead to the above $\d H_7$.

\end{document}